\documentclass[a4paper,twocolumn]{article}

\usepackage[english]{babel}

\usepackage[margin=1.9cm]{geometry}
\usepackage{amsmath}
\usepackage{amsfonts}
\usepackage{subfigure}

\usepackage[title,titletoc,toc]{appendix}

\usepackage{tikz}
\usepackage{pgf-umlsd}
\usepackage{pgfplots}
\usepackage{textcomp}
\usepackage{mathtools}
\usepackage[mathcal]{euscript}
\usepackage{array}
\usepackage{lipsum}
\usepackage{cuted}
\usepackage{flushend}
\usepackage{placeins}
\usepackage{float}
\usepackage[hidelinks]{hyperref}
\usepackage{wrapfig}

\usetikzlibrary{arrows}
\usetikzlibrary{shapes.geometric, shadows}
\usetikzlibrary{patterns}
\usetikzlibrary{positioning}
\usetikzlibrary{external}
\usetikzlibrary{decorations.text}
\usetikzlibrary{plotmarks}
\pgfplotsset{compat=newest}

\usepgfplotslibrary{dateplot}

\newlength{\parin}
\newenvironment{widetext}
{
    \begin{strip}
    \rule{\dimexpr(0.5\textwidth-0.5\columnsep-0.4pt)}{0.4pt}\rule{0.4pt}{6pt}
    \vspace*{-1.5mm}\par\parindent\parin
}
{
    \end{strip}
}

\title{\textbf{A Bayesian Approach to Identify Bitcoin Users}}

\newcolumntype{C}[1]{>{\centering\let\newline\\\arraybackslash\hspace{0pt}}m{#1}}

\author{
\begin {tabular} {c@{\hskip 11mm}c@{\hskip 11mm}c@{\hskip 11mm}c} \centering
P\'eter L. Juh\'asz$^{1,2}$ & J\'ozsef St\'eger$^1$ & D\'aniel Kondor$^{1,3}$ & G\'abor Vattay$^1$ \\
\scriptsize {peter.laszlo.juhasz@ericsson.com} & \scriptsize {steger@complex.elte.hu} & \scriptsize {dkondor@mit.edu} & \scriptsize {vattay@complex.elte.hu}
\end {tabular}
\and
\begin {tabular} {rl} \centering ~ & ~ \\
\footnotesize {1} & \footnotesize {Department of Physics of Complex Systems, E\"otv\"os Lor\'and University} \\[-1.3mm]
~ & \footnotesize {1/A P\'azm\'any P\'eter s\'et\'any, Budapest, H-1117 Hungary} \\
\footnotesize {2} & \footnotesize {Business Unit IT \& Cloud Products, Ericsson Telecommunications Hungary} \\[-1.3mm]
~ & \footnotesize {4 Irinyi J\'ozsef utca, Budapest, H-1117 Hungary} \\
\footnotesize {3} & \footnotesize {SENSEable City Laboratory, Massachusetts Institute of Technology} \\[-1.3mm]
~ & \footnotesize {77 Massachusetts Avenue, Cambridge, MA 02139 USA}
\end {tabular}
}

\date{\vspace{-4ex}}

\mathtoolsset {showonlyrefs}

\tikzstyle {gep}   = [circle, text centered, draw = black, fill = white, line width = .5mm, text width = 1cm, font = \large]
\tikzstyle {tx}    = [rectangle, minimum width = 3cm, minimum height = 1cm, draw = black, fill = white, line width = .15mm, text width = 3.7cm]
\tikzstyle {arrow} = [-triangle 45, line width = .15mm, postaction = {draw, line width = .15mm, shorten > = 1mm, -}]

\tikzstyle {element}       = [rectangle, rounded corners, minimum width=3cm, minimum height=1.8cm, text centered, draw=black, fill=white, line width=.15mm, text width=3cm]
\tikzstyle {jelentektelen} = [rectangle, dashed,          minimum width=1cm,   minimum height=1cm,   text centered, draw=black, fill=white, line width=.15mm, text width=2.3cm, font = \small]
\tikzstyle {arrowf}        = [-triangle 45, line width = .15mm, postaction = {draw, line width = .15mm, shorten > = 1mm, -}]
\tikzstyle {arrowj}        = [-triangle 45, line width = .1mm,  postaction = {draw, line width = .1mm , shorten > = 1mm, -}]

\begin{document}

\newcommand {\ud  } {\mathrm {d}  }
\newcommand {\expe} {\mathrm {exp}}

\maketitle

\subsubsection* {Abstract}

Bitcoin is a digital currency and electronic payment system operating over a peer-to-peer network on the Internet.
One of its most important properties is the high level of anonymity it provides for its users.
The users are identified by their Bitcoin addresses, which are random strings in the public records of transactions, the \emph {blockchain}.
When a user initiates a Bitcoin-transaction, his Bitcoin client program relays messages to other clients through the Bitcoin network.
Monitoring the propagation of these messages and analyzing them carefully reveal hidden relations.
In this paper, we develop a mathematical model using a probabilistic approach to link Bitcoin addresses and transactions to the originator IP address.
To utilize our model, we carried out experiments by installing more than a hundred modified Bitcoin clients distributed in the network to observe as many messages as possible. 
During a two month observation period we were able to identify several thousand Bitcoin clients and bind their transactions to geographical locations.

\subsubsection* {Keywords}

bitcoin; anonymity; peer-to-peer network; bayes classifier; bitcoin protocol; identification

\section {Introduction}

Bitcoin is the first widely used digital currency, developed by Satoshi Nakamoto after the beginning of the financial crisis in 2009~\cite {nakamoto}.
A distinctive feature of Bitcoin is that there is no central authority overseeing transactions, users are connected via a peer-to-peer network where they announce any transaction they wish to make.
Transactions can then be validated by anyone using the publicly available list of transactions, the \emph {blockchain}, which is in turn generated in a proof-of-work system.
Cheating (e.g.~including invalid transactions in the blockchain) thus would require one entity to control more than $50\%$ of the computing power that users dedicate to generating the blockchain.
In accordance with the decentralized nature of the system, the specifications of the network protocol is publicly available, while several open-source client programs implementing the protocol exist~\cite {btc_repo}.

One of the key characteristics of Bitcoin is the high amount of anonymity it provides for its users~\cite {anonymity}.
Although one can learn the details of the transactions via the blockchain, it is still unknown who the users initiating those transactions are.
This is possible since as there is no authority overseeing the operation of the system, users do not need to provide any form of identification to join; anyone with an Internet connection can download a client program, which then allows them to generate any number of Bitcoin addresses that they can use in the transactions to send or receive Bitcoins.
This results in that the identity of Bitcoin users is only revealed if they publish their Bitcoin address or this information is intercepted in some way outside the Bitcoin system.
While anonymity is not among the main design goals of the Bitcoin system~\cite {anonymity}, Bitcoin is widely considered as a highly anonymous way of performing financial transactions and is often utilized for illegal uncontrolled payments~\cite {ujhet}, along legal uses where the involved parties do not wish to disclose their identities to controlling entities in the traditional financial system, e.g.~banks or governments.

In the paper, we present a probabilistic model based on the information propagating over the Bitcoin network, which gives the possibility of identifying the users initiating the transactions.
In this case, identification means binding the transactions to the IP addresses where they were created.

The basic idea consists of two main steps.
First, the probability is determined for each transaction that a specific client (identified by its IP address) created it.
Assuming that the creator of the transaction controls the Bitcoin addresses from which money is sent in it, this step then results in possible IP address -- Bitcoin address pairings.
Next the most likely Bitcoin address -- client pairings are identified by combining the probabilities in the list of pairings compiled in the previous step.
This is further elaborated by grouping Bitcoin addresses that belong to the same user with high probability based on the transaction network.
Finally, the geographical localization of the IP addresses opens the door for a large scale analysis of the distribution and flow of Bitcoin.

The rest of the paper is structured as follows.
Section~\ref {sec_main_char} discusses the relevant characteristics of Bitcoin and provides the necessary background for the further sections.
In section~\ref {sec_bayesian}, the mathematical model used for the deanonymization is explained.
The data collection is described in section~\ref {sec_datacoll}.
Section~\ref {sec_results} presents the results of the application of the model.
Finally the method described in this study is compared to the related works of the topic in section~\ref {sec_related}.

\section{The Main Characteristics of Bitcoin}
\label{sec_main_char}

In order to use Bitcoin one has to connect to the Bitcoin network using an open-source \emph {client program}~\cite {btc_repo}.
In this work, we concentrate on the Bitcoin Core client~\cite {btc_repo}, whose source code we inspected and modified for the purpose of data collection.
By default, this client establishes eight connections to other clients.
If there is a link between two clients, they are \emph {connected}.
Clients exchange information of different types, e.g.~the transactions they know about, their state, cryptographic signatures and others through the network.
This is necessary for the validation of the transactions as it is done by the entire network.

In case of Bitcoin transactions, \emph {Bitcoin addresses} play similar role as the bank account numbers in regular currency transactions. 
However, there are two major differences:
\begin {samepage} \begin {itemize}
\item {each user may have as many Bitcoin addresses as they would like to}
\item {and multiple source and destination Bitcoin addresses can be involved in a single transaction.}
\end {itemize} \end {samepage}

In case of the Bitcoin Core client program that was in operation at the time of the measurement, when a user initiates a transaction, the client program (the \emph {originator}) relays a message to a randomly chosen connected client in every $100\,ms$ time interval.
This method is referred to as \emph {trickling}, and its goal is to hide the source of the transaction.
The clients receiving this message (which are not the originators of the transaction) use a slightly more complex algorithm to further send the information.
Besides trickling, they also relay the message to the other clients with a probability of $1/4$ (in every $100\,ms$).
We expect that other types of clients apply the same mechanisms to protect the privacy of the users.

We note that as of today, the previously described mechanism for relaying transactions has been changed in the case of the Bitcoin Core client.
Currently every client maintains a queue for the messages to be relayed for each connected clients and relays them according to a Poisson process.
The parameter of the process is $5\,sec$ for incoming connections and $2.5\,sec$ for outgoing connections.
In this work, we consider the previously described method which was in use during the time of our data collection; we believe that our model could be used for the latter case as well with minor modifications.

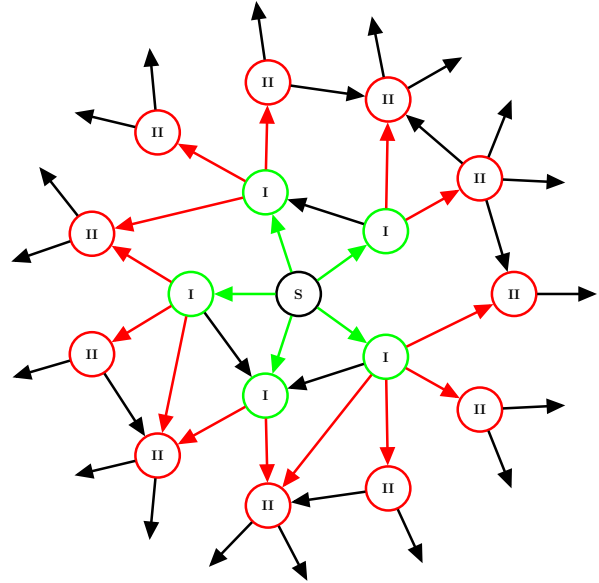
\begin {figure} [ht] \centering \resizebox {.45\textwidth} {!} {\begin {tikzpicture} [scale=0.6, every node/.style={scale=0.65}, node distance=0cm]
\node (s) [gep]                                                 {\textbf {S}};
\node (a) [gep, draw=green, xshift=-3.25cm  , yshift=0cm      ] {\textbf {I}};
\node (b) [gep, draw=green, xshift=-1.0043cm, yshift=-3.0909cm] {\textbf {I}};
\node (c) [gep, draw=green, xshift=2.6293cm , yshift=-1.9103cm] {\textbf {I}};
\node (d) [gep, draw=green, xshift=2.6293cm , yshift=1.9103cm ] {\textbf {I}};
\node (e) [gep, draw=green, xshift=-1.0043cm, yshift=3.0909cm ] {\textbf {I}};

\draw [->,-triangle 45, line width=.5mm, green, postaction] (s) -- (a);
\draw [->,-triangle 45, line width=.5mm, green, postaction] (s) -- (b);
\draw [->,-triangle 45, line width=.5mm, green, postaction] (s) -- (c);
\draw [->,-triangle 45, line width=.5mm, green, postaction] (s) -- (d);
\draw [->,-triangle 45, line width=.5mm, green, postaction] (s) -- (e);

\draw [->,-triangle 45, line width=.5mm,        postaction] (a) -- (b);
\draw [->,-triangle 45, line width=.5mm,        postaction] (c) -- (b);
\draw [->,-triangle 45, line width=.5mm,        postaction] (d) -- (e);

\node (f) [gep, draw=red, xshift=6.5cm,     yshift=0cm      ] {\textbf {II}};
\node (g) [gep, draw=red, xshift=5.4681cm,  yshift=3.5142cm ] {\textbf {II}};
\node (h) [gep, draw=red, xshift=2.7002cm,  yshift=5.9126cm ] {\textbf {II}};
\node (i) [gep, draw=red, xshift=-0.9251cm, yshift=6.4338cm ] {\textbf {II}};
\node (j) [gep, draw=red, xshift=-4.2566cm, yshift=4.9124cm ] {\textbf {II}};
\node (k) [gep, draw=red, xshift=-6.2367cm, yshift=1.8312cm ] {\textbf {II}};
\node (l) [gep, draw=red, xshift=-6.2367cm, yshift=-1.8312cm] {\textbf {II}};
\node (m) [gep, draw=red, xshift=-4.2566cm, yshift=-4.9124cm] {\textbf {II}};
\node (n) [gep, draw=red, xshift=-0.9250cm, yshift=-6.4338cm] {\textbf {II}};
\node (o) [gep, draw=red, xshift=2.7002cm,  yshift=-5.9126cm] {\textbf {II}};
\node (p) [gep, draw=red, xshift=5.4682cm,  yshift=-3.5141cm] {\textbf {II}};

\draw [->,-triangle 45, line width=.5mm, red, postaction] (a) -- (k);
\draw [->,-triangle 45, line width=.5mm, red, postaction] (a) -- (l);
\draw [->,-triangle 45, line width=.5mm, red, postaction] (a) -- (m);
\draw [->,-triangle 45, line width=.5mm, red, postaction] (b) -- (m);
\draw [->,-triangle 45, line width=.5mm, red, postaction] (b) -- (n);
\draw [->,-triangle 45, line width=.5mm, red, postaction] (c) -- (n);
\draw [->,-triangle 45, line width=.5mm, red, postaction] (c) -- (o);
\draw [->,-triangle 45, line width=.5mm, red, postaction] (c) -- (p);
\draw [->,-triangle 45, line width=.5mm, red, postaction] (c) -- (f);
\draw [->,-triangle 45, line width=.5mm, red, postaction] (d) -- (g);
\draw [->,-triangle 45, line width=.5mm, red, postaction] (d) -- (h);
\draw [->,-triangle 45, line width=.5mm, red, postaction] (e) -- (i);
\draw [->,-triangle 45, line width=.5mm, red, postaction] (e) -- (j);
\draw [->,-triangle 45, line width=.5mm, red, postaction] (e) -- (k);

\draw [->,-triangle 45, line width=.5mm,      postaction] (l) -- (m);
\draw [->,-triangle 45, line width=.5mm,      postaction] (o) -- (n);
\draw [->,-triangle 45, line width=.5mm,      postaction] (g) -- (f);
\draw [->,-triangle 45, line width=.5mm,      postaction] (g) -- (h);
\draw [->,-triangle 45, line width=.5mm,      postaction] (i) -- (h);

\draw [->,-triangle 45, line width=.5mm,      postaction] (f) -- (9.7500,0);
\draw [->,-triangle 45, line width=.5mm,      postaction] (i) -- (-1.3877,9.6507);
\draw [->,-triangle 45, line width=.5mm,      postaction] (o) -- (4.0503,-8.8689);
\draw [->,-triangle 45, line width=.5mm,      postaction] (l) -- (-9.3551,-2.7468);

\draw [->,-triangle 45, line width=.5mm,      postaction] (g) -- (6.9573,6.4029);
\draw [->,-triangle 45, line width=.5mm,      postaction] (g) -- (8.7144,3.6689);
\draw [->,-triangle 45, line width=.5mm,      postaction] (h) -- (2.3913,9.1479);
\draw [->,-triangle 45, line width=.5mm,      postaction] (h) -- (5.3476,7.7978);
\draw [->,-triangle 45, line width=.5mm,      postaction] (j) -- (-7.3279,5.9754);
\draw [->,-triangle 45, line width=.5mm,      postaction] (j) -- (-4.8717,8.1037);
\draw [->,-triangle 45, line width=.5mm,      postaction] (k) -- (-9.3951,1.0650);
\draw [->,-triangle 45, line width=.5mm,      postaction] (k) -- (-8.4791,4.1833);
\draw [->,-triangle 45, line width=.5mm,      postaction] (m) -- (-4.8717,-8.1037);
\draw [->,-triangle 45, line width=.5mm,      postaction] (m) -- (-7.3279,-5.9754);
\draw [->,-triangle 45, line width=.5mm,      postaction] (n) -- (0.2829,-9.4500);
\draw [->,-triangle 45, line width=.5mm,      postaction] (n) -- (-2.9340,-8.9885);
\draw [->,-triangle 45, line width=.5mm,      postaction] (p) -- (8.7145,-3.6687);
\draw [->,-triangle 45, line width=.5mm,      postaction] (p) -- (6.9575,-6.4028);

\end {tikzpicture}} \caption {A new transaction is initiated by the client ``S''. At first it informs the clients denoted by ``I'' (they are informed directly from the originator, so only the trickling method is used for the relay). Then, these clients relay the transaction further -- among possibly other I type clients -- to the ones denoted by ``II''.} \label {propagation} \end {figure}

In accordance with the previously described methodology, the network relies on clients relaying transactions to have them spread throughout the entire network.
As a consequence an arbitrarily chosen client is not necessarily directly informed about the transaction by the originator (see Figure~\ref {propagation})

As no state, bank, institute or organization controls or ensures the validity of Bitcoin transactions, cryptographic methods are used by the whole Bitcoin community for this purpose.
The security of Bitcoin is based on the blockchain.
In this study the source Bitcoin addresses, the destination Bitcoin addresses, the timestamps and the transferred volume of Bitcoin is extracted from the blockchain for each transaction.

If the owners of the Bitcoin addresses were known, the blockchain would reveal all of the transactions of each Bitcoin user.
The open nature of the system mitigates this concern, as anyone can generate any number of Bitcoin addresses without having to reveal their identity.
Nevertheless, if a Bitcoin address can be linked to someone (either because they share it in order to receive Bitcoins or by any other method), the transaction history of that Bitcoin address can be trivially retrieved from the blockchain.
Thus, keeping the association between Bitcoin addresses and real identities in secret is crucial for users who wish to maintain their privacy.

\section {A Bayesian Method for the Identification of Bitcoin Users}
\label {sec_bayesian}

In this section we present the methodology to assign probabilities to the distinct IP address -- user pairings, which consists of three main steps.

\begin {figure} [!htb] \centering \resizebox {.45\textwidth} {!} {\begin {tikzpicture} [node distance=0cm]
\node (IP)     [element]                                                 {\small {Identify\\the Transactions' Possible Originator IP Addresses}};
\node (felh)   [element, below=of IP, xshift=0cm,yshift=-1cm]            {\small {Group Bitcoin Addresses}};
\node (felhIP) [element, xshift=5cm,yshift=-1.4cm]                     {\small {Assign\\IP Addresses to Users}};
\node (db)     [jelentektelen, below=of felh, xshift=0cm, yshift=-0.7cm] {\small {Blockchain}};
\draw [arrowf] (IP)   to [out=0,in=180]                   node [anchor=south,midway,yshift=.1cm] {} (felhIP);
\draw [arrowf] (felh) to [out=0,in=180]                   node [anchor=south,midway,yshift=.1cm] {} (felhIP);
\draw [arrowj] (db)   to [out=90,in=-90, line width=.3mm] node [anchor=south,midway,yshift=.1cm] {} (felh);
\end {tikzpicture}} \caption {Main Steps} \label {mainsteps} \end {figure}
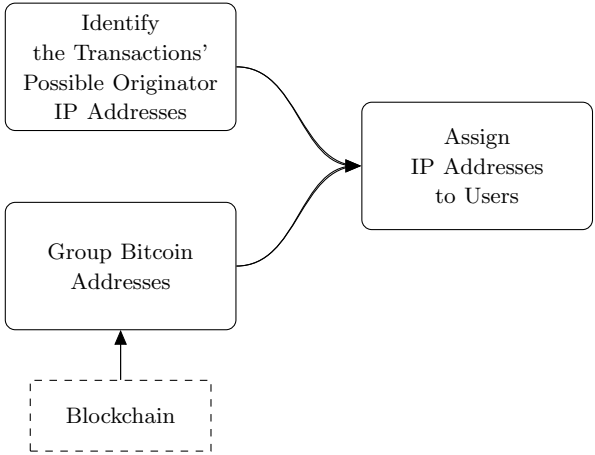

An overview of the process is illustrated in Figure~\ref {mainsteps}.

First, the propagating messages are observed and recorded by several \emph {monitoring clients} in order to cover as great part of the network as possible.
For each transaction, monitoring clients record the list of clients who relayed the transaction in the \emph {first time segment} (see the definition in the next subsection).
They are the possible originators of the transaction.
After some theoretical considerations, we assign probabilities to each client that show the probability of them being the originator, separately for each transaction that we recorded.

Next, the blockchain is used to group the Bitcoin addresses owned by the same user.
Additionally, the blockchain also enables to calculate the balances of the users for further analysis.

Last, by having possibly several transactions of the same Bitcoin address and the grouping of Bitcoin addresses by user allows us to combine measurements from multiple transactions to identify users with higher confidence.
By combining the probabilities from the first step, the users (and their balances) are paired with the clients that are most likely the originators of their transactions.
The clients can be geographically localized through their IP addresses, which allows the determination of the geographical distribution and flow of Bitcoins.

\subsubsection* {Step 1: Individual Probabilities}

Let us consider a single transaction observed by one \emph {monitoring client}.

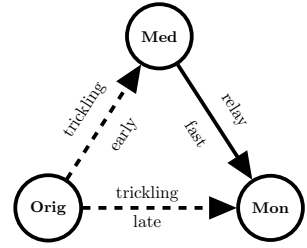
\begin {figure} [ht] \centering
\begin {tikzpicture} [scale=0.6, every node/.style={scale=0.65}, node distance=0cm]
\node (o) [gep                             ] {\small {\textbf {Orig}}};
\node (m) [gep, xshift=4.5cm , yshift=0cm  ] {\small {\textbf {Mon}}};
\node (c) [gep, xshift=2.25cm, yshift=3.5cm] {\small {\textbf {Med}}};
\draw [->,-triangle 45, line width=.5mm, postaction, dashed] (o) -- (m) node [above, midway, sloped, xshift=-0.25cm ] {trickling} node [below, midway, sloped, xshift=-0.25cm ] {late};
\draw [->,-triangle 45, line width=.5mm, postaction, dashed] (o) -- (c) node [above, midway, sloped, outer sep=0.2cm] {trickling} node [below, midway, sloped, outer sep=0.2cm] {early};
\draw [->,-triangle 45, line width=.5mm, postaction        ] (c) -- (m) node [above, midway, sloped, outer sep=0.2cm] {relay}     node [below, midway, sloped, outer sep=0.2cm] {fast};
\end {tikzpicture} \caption {The message can be routed faster from the originator (Orig) to the monitoring client (Mon) through a mediator (Med) in the shown scenario.} \label {mediator} \end {figure}

A monitoring client connected to the originator does not necessarily receive the message from the originator first, because in some cases it can be relayed faster through a mediator client (Figure~\ref {mediator}).

One iteration of sending messages happens every $100\,ms$ time interval.
Let us first calculate the probability that the originator relays the message to the monitoring client in one specific iteration.
If the originator has $c_\textrm {orig}$ clients connected to it (among which one is the monitoring client), then in every iteration there is $1/c_\textrm {orig}$ probability, that it relays to the monitoring client.
\begin {equation} \mathbb {P}_1^\textrm {orig} = \frac {1} {c_\textrm {orig}} \end {equation}
In case of the mediator client, it relays the transaction to the monitor client with a probability of $1/c_\textrm {med}$ because of trickling, and it relays with a probability of $1/4$ if the other mechanism is used.
\begin {equation} \mathbb {P}_1^\textrm {med} = \frac {1} {c_\textrm {med}} + \frac {1} {4} \left( 1 - \frac {1} {c_\textrm {med}} \right) = \frac {1} {4} - \frac {3} {4 c_\textrm {med}} \end {equation}

The probability that a specific client relays the transaction in the $k$-th iteration, follows a geometric distribution.
\begin {equation} \begin {split} \mathbb {P}_k^\textrm {orig} &= \mathbb {P}_1^\textrm {orig} \left( 1 - \mathbb {P}_1^\textrm {orig} \right)^{k - 1} \\ \mathbb {P}_k^\textrm {med} &= \mathbb {P}_1^\textrm {med} \left( 1 - \mathbb {P}_1^\textrm {med} \right)^{k - 1} \end {split} \end {equation}

Let us consider the route on Figure \ref {mediator} when the originator sends the message to the mediator, and then the mediator further relays it to the monitoring client.
To calculate the distribution of the iterations for the route through the mediator client, the sum of the two random variables has to be considered.
This can be derived from the discrete convolution of the above two distributions.
\begin {equation} \begin {split} &\mathbb {P}^\textrm {orig + med}_k = \sum_{i = 1}^{k - 1} \left( \mathbb {P}_i^\textrm {orig} + \mathbb {P}_{k-i}^\textrm {med} \right) = \\ &= \frac {\mathbb {P}_1^\textrm {orig} \, \mathbb {P}_1^\textrm {med} \left( 1 - \mathbb {P}_1^\textrm {med} \right)^{k - 1}} {1 - \mathbb {P}_1^\textrm {orig}} \sum_{i = 1}^{k - 1} \left( \frac {1 - \mathbb {P}_1^\textrm {orig}} {1 - \mathbb {P}_1^\textrm {med}} \right)^i = \\ &= \frac {\mathbb {P}_1^\textrm {orig} \, \mathbb {P}_1^\textrm {med}} {\mathbb {P}_1^\textrm {med} - \mathbb {P}_1^\textrm {orig}} \left[ \left( 1 - \mathbb {P}_1^\textrm {orig} \right)^{k - 1} - \left( 1 - \mathbb {P}_1^\textrm {med} \right)^{k - 1} \right] \end {split} \end {equation}

As every ordinary client initiates $8$ outgoing connections when connecting to the network, the number of connections is estimated to be $16$ (taking into account the incoming connections as well).

If the two routes on Figure \ref {mediator} are considered independent from each other, the probability of the direct route being shorter (i.e.~requires less iterations) is $0.6334$.\footnote {Here we have not taken into account the network delay, and that multiple indirect route can exist from the originator to the monitoring client possibly consisting of more steps. With this model however, we can approximate the probabilities.}
The goal is to determine a time frame that the monitoring client has to wait after receiving the message first until surely receiving it directly from the originator, if they are directly connected.
If this waiting time is defined to be $2\,sec$, the above model gives a probability of $0.8071$ for the direct route taking less iterations.

Since the time of our data collection, the trickling mechanism has been changed in the case of the Bitcoin Core client, so that relaying can be described by Poisson processes.
In this case a similar calculation could be utilized but the waiting time will need to be adjusted to a value which maintains a high probability for the direct route.
However, this does not change the further steps of our model except from the derivation of the above probabilities.

To successfully relay a transaction to another client, three messages have to be exchanged.
First, the sender informs the receiver about the transactions it knows about (``INV'' message).
Then the receiver asks for the new, unknown transactions in the answer (``GETDATA'' message).
Finally, the actual information is sent to the receiver (``TX'' message).
As three messages need to be exchanged sequentially, the delay of the network plays an important role in the message propagation.
The more mediator clients are involved in the transmission of the transaction, the longer the time it takes for the message to get to the monitoring client from the originator.

If we take into account the network delay, and that the above described ``worst case scenario'' (i.e.~ that we are connected to the originator, and an indirect route consisting of one mediator exists) is unlikely, we can \emph {neglect} the probability that a message is received from an indirect route earlier than two seconds before receiving it from the originator.
This assumption is experimentally verified in~\cite {ujot}.

We call this time interval the \emph {first time segment} of the transaction and denote it by $t_1 = 2\,sec$.
If the monitoring client is not connected directly to the originator, it will only receive the transaction via possibly multiple indirect routes.
Nevertheless, it will be true with high probability that connected clients that do not belong to this first time segment are not the originators of the transaction.
We then proceed with this assumption to estimate the probabilities of a client being the originator of the transaction based on each received transaction.

As of today, this mechanism has changed, so that the relay can be described by Poisson processes.
In this case a similar calculation can be utilized, except from the fact that the probability of not belonging to the first time segment can not be neglected.

From the perspective of a monitoring client, the other Bitcoin clients can be classified to sets based on each transaction according to Figure~\ref {individual}.
Some of the Bitcoin clients relay the message to it in the first time segment.
This constitutes a subset of the Bitcoin clients to which the monitoring client is connected to at the time of the transaction.
Only active Bitcoin clients are connected to the network, but not all of the clients are working at the examined moment.
\def \firstcircle  {(0,0) circle (3cm)}
\def \secondcircle {(0,0) circle (2cm)}
\def \thirdcircle  {(0,0) circle (1cm)}
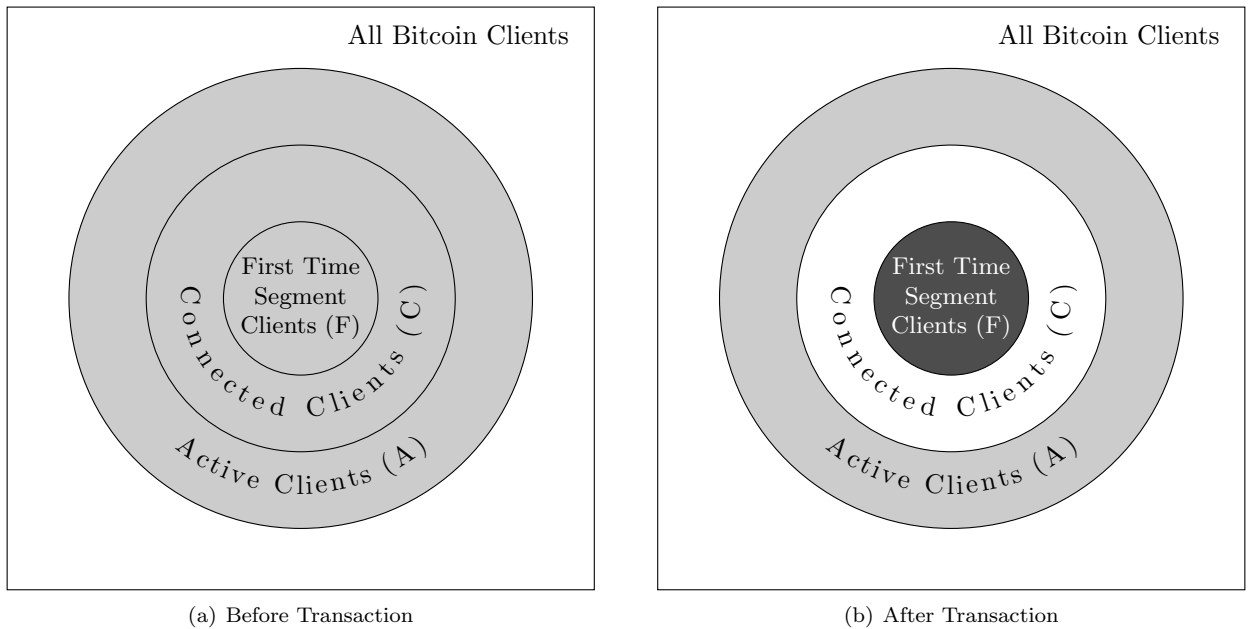
\begin {figure*} [!t] \centering
\subfigure [Before Transaction] {\resizebox {0.45\textwidth} {!} {\begin {tikzpicture}
    \draw (-3.8,-3.8) rectangle (3.8,3.8) node [pos=.9, xshift=-1cm, yshift=.4cm] {All Bitcoin Clients};
    \draw [fill = black!20!white] \firstcircle node[above = 2.2cm] {};
    \draw [cm = {cos(-90), -sin(-90), sin(-90), cos(-90), (0,0)}, postaction = {decorate}, decoration = {raise = -3.6ex, text along path, text align = center, text = {Active Clients (A)}}] \secondcircle node [above = 1.2cm] {};
    \draw [cm = {cos(-90), -sin(-90), sin(-90), cos(-90), (0,0)}, postaction = {decorate}, decoration = {raise = -3.6ex, text along path, text align = center, text = {Connected Clients (C)}}] \thirdcircle node [align = center, font = \small] {First Time\\Segment\\Clients (F)}; \end {tikzpicture}}}
\qquad
\subfigure [After Transaction] {\resizebox {0.45\textwidth} {!} {\begin {tikzpicture}
    \draw (-3.8,-3.8) rectangle (3.8,3.8) node [pos=.9, xshift=-1cm, yshift=.4cm] {All Bitcoin Clients};
    \draw [fill = black!20!white] \firstcircle node[above = 2.2cm] {};
    \draw [fill = white, cm = {cos(-90), -sin(-90), sin(-90), cos(-90), (0,0)}, postaction = {decorate}, decoration = {raise = -3.6ex, text along path, text align = center, text = {Active Clients (A)}}] \secondcircle node [above = 1.2cm] {};
    \draw [fill = black!70!white,
cm = {cos(-90), -sin(-90), sin(-90), cos(-90), (0,0)}, postaction = {decorate}, decoration = {raise = -3.6ex, text along path, text align = center, text = {Connected Clients (C)}}] \thirdcircle node [align = center,  font = \small, text = white] {First Time\\Segment\\Clients (F)}; \end {tikzpicture}}}
\caption {Relation of the different sets of clients. The darker the subset is, the higher the probability of a client in that subset is the originator of the transaction.} \label {individual} \end {figure*}

Before the transaction, no information is known, thus the best estimate we can make is that each Bitcoin client has equal probability of being the originator of the transaction, resulting in a uniform probability distribution among the active clients (left side of Figure~\ref {individual}).
After the transaction, each Bitcoin client in the first time segment can be either the real originator of the transaction or a client relaying it (via several hops).
Furthermore, the real originator can also be among the rest of the network, not connected to our monitoring client.
On the other hand, based on the previous arguments, we presume that clients not relaying the transaction in the first time segment are certainly not the originators of the transaction.
Thus, the probability of the first time segment clients increases while the connected clients not belonging to the first time segment will have zero probability (right side of Figure~\ref {individual}).
Still nothing is known about the clients not connected to the monitoring client, therefore their probabilities will not change.
Also, clients belonging to the same subsets can not be distinguished.

Let us calculate the probabilities of being the originator for clients in each set.
The Roman font type notations of Figure~\ref {individual} are used for the sets.
The number of elements in the sets is denoted by $\left| \, \cdot \, \right|$.
$\mathcal {C}$ denotes that the monitoring client is connected to the originator of the transaction, $\mathcal {O}$ denotes that the originator relays the message in the first time segment to the monitoring client and $\mathcal {F}$ means that a randomly chosen client from the first time segment is actually the originator of the transaction.
Using these notations, we have that
\begin {equation} \mathbb {P} \left( \mathcal {C} \right) = \frac {\left| C \right|} {\left| A \right|} \end {equation}
as inactive clients can not be the originator of the transaction.
If the monitoring client is connected to the originator, it is going to inform the monitoring client in the first time segment.
At this time all of the first time segment clients have the same probability of being the originator.
\begin {equation} \mathbb {P} \left( \mathcal {O} | \mathcal {C} \right) = 1 \qquad \mathbb {P} \left( \mathcal {F} | \mathcal {C} \right) = \frac {1} {\left| F \right|} \end {equation}
Let us apply the law of total probability for $\mathbb {P} \left( \mathcal {F} \right)$.
\begin {equation} \mathbb {P} \left( \mathcal {F} \right) = \mathbb {P} \left( \mathcal {F} | \mathcal {C} \right) \cdot \mathbb {P} \left( \mathcal {C} \right) + \mathbb {P} \left( \mathcal {F} | \overline {\mathcal {C}} \right) \cdot \mathbb {P} \left( \overline {\mathcal {C}} \right) = \frac {\left| C \right|} {\left| A \right| \, \left| F \right|} \end {equation}
where we exploited that a client can not send any messages in the first time segment if it is not at all connected to the monitoring client: $\mathbb {P} \left( \mathcal {F} | \overline {\mathcal {C}} \right) = 0$.

The above formula gives the probability assigned to the first time segment clients.
The connected clients not belonging to the first time segment have zero probability.
The rest of the active clients has the same $1 / \left| A \right|$ probability.
We note that these probabilities still sum up to $1$: probabilities among the connected clients were ``redistributed'' according to whether they belong to the first time segment.

So far we only considered one monitoring client.
If there are more monitoring clients the above mentioned sets are defined separately for each of them, and then the union of the corresponding sets is determined, i.e.~$F(tx_i) = \cup_{j=1}^N F_j(tx_i)$ and $C(tx_i) = \cup_{j=1}^N C_j(tx_i)$ for $N$ monitoring clients, where the subscripts denote the corresponding sets as observed for transaction $tx_i$ by the $j$th monitoring client.
Using this method, monitoring Bitcoin clients do not need to be synchronized in time.
If time synchronization among monitoring clients was achieved, we could further limit the $F$ set of first time segment clients to those that broadcast the transaction in $t_1$ time after \emph {any} of our monitoring clients first received the transaction.
In our experiments, achieving reliable time synchronization was not possible, so the union of sets was used as described.
We note that the set of active clients at a given time ($A$) is not straightforward to estimate even with a large number of monitoring clients.
To do that, we would need to perform an active network discovery over the peer-to-peer network of Bitcoin clients.
Instead of implementing this functionality ourselves, we relied on the Bitnodes.io database~\cite {getaddr}, which provides the estimated number of active Bitcoin clients as a function of time (i.e.~$|A|$).
The actual set is not required for the calculations, only the size of the set at the time of the transactions is considered.

\subsubsection* {Step 2: Grouping the Transactions Belonging to the Same User}

The next task is to group the Bitcoin addresses according to the users they are owned by.
After this, every transaction can be assigned to the users by looking at the source Bitcoin addresses of the transaction.

To group addresses, we exploit that Bitcoin addresses appearing on the input side of the same transaction typically belong to the same user.
This assumption is employed widely in the literature as well~\cite {ujketto,ujhat,ujharom,ujegy}.

\begin {figure*} [htb] \centering \resizebox {.7\textwidth} {!} {\begin {tikzpicture} [node distance=0cm]
\node (txa) [tx] {\small {\textbf {Transaction 1}} \\
\footnotesize {1BsWmvFJ4oqgVVvs1cFE} \\
\footnotesize {1BdVGd582jsQYBeLYqtg} \\
\small {\textbf {\textcolor {red} {16LAiE7S3pfA53U5E}}} \\
\footnotesize {1M6xUmHyKpvWtcusg4k3}
};
\node (txb) [tx, below=of txa, xshift=0cm, yshift=-0.5cm] {\small {\textbf {Transaction 2}} \\
\small {\textbf {\textcolor {red} {16LAiE7S3pfA53U5E}}} \\
\footnotesize {19dqKRxDpsfRmQnRAd22} \\
\footnotesize {1C3xKJaMG1C2yQXfmfG}
};
\node (txc) [tx, below=of txb, xshift=0cm, yshift=-0.5cm] {\small {\textbf {Transaction 3}} \\
\footnotesize {1Nw39NH5eRRtmqcPwRTc} \\
\footnotesize {1EMcpdDbY6W53mDWou} \\
\footnotesize {1DbD7zFjYSQRiATc8dHx}
};
\node (fa) [tx, right=of txa, rounded corners, xshift=3cm, yshift=-1.3cm] {\small {\textbf {User 1}} \\
\footnotesize {1BsWmvFJ4oqgVVvs1cFE} \\
\footnotesize {1BdVGd582jsQYBeLYqtg} \\
\small {\textbf {\textcolor {red} {16LAiE7S3pfA53U5E}}} \\
\footnotesize {1M6xUmHyKpvWtcusg4k3} \\
\footnotesize {19dqKRxDpsfRmQnRAd22} \\
\footnotesize {1C3xKJaMG1C2yQXfmfG}
};
\node (fb) [tx, below=of fa, rounded corners, xshift=0cm, yshift=-1cm] {\small {\textbf {User 2}} \\
\footnotesize {1Nw39NH5eRRtmqcPwRTc} \\
\footnotesize {1EMcpdDbY6W53mDWou} \\
\footnotesize {1DbD7zFjYSQRiATc8dHx}
};
\draw [arrow] (txa) to [out=0,in=180] node [anchor=south,midway,yshift=.1cm] {} (fa);
\draw [arrow] (txb) to [out=0,in=180] node [anchor=south,midway,yshift=.1cm] {} (fa);
\draw [arrow] (txc) to [out=0,in=180] node [anchor=south,midway,yshift=.1cm] {} (fb);
  \end {tikzpicture}} \caption {Grouping of Bitcoin addresses: the left side shows three transactions and the input Bitcoin addresses of these transactions, while the right side indicates how these Bitcoin addresses are grouped.} \label {group} \end {figure*}
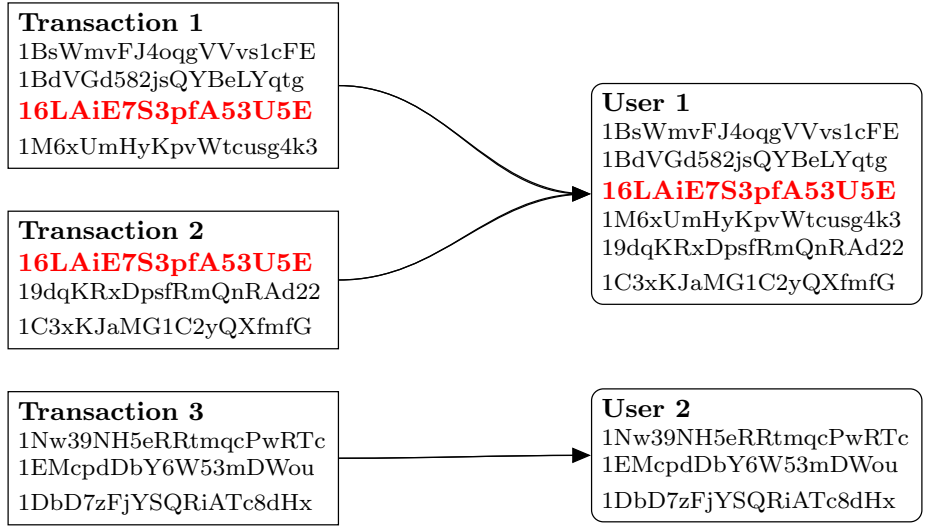

This can be used for grouping individual Bitcoin addresses. The process is demonstrated in Figure~\ref {group}.
The left side of the figure shows the transactions and the input Bitcoin addresses where the Bitcoins are sent from.
These Bitcoin addresses belong to the same user.
When a Bitcoin address appears in different transactions (marked red and bold), all Bitcoin addresses can be merged and assigned to the same user.
Although Bitcoin users are encouraged to generate new Bitcoin addresses after every transaction they make, so that the above grouping is less efficient~\cite {mixcoin}, most of the users do not follow this guideline~\cite {koshyphd,fistful}.

The transactions belong to the user that owns its input Bitcoin address(es).

\subsubsection* {Step 3: Combining Probabilities -- Naive Bayes Classification}

From the message propagation it can be determined how likely the clients are the originators of the transactions.
So far we considered the transactions independently from each other.

According to our assumptions, the transactions belonging to a single user were created by a few originator clients.
This means that these transactions provide probabilities for the same set of originator clients.
The originator clients can be identified more efficiently by combining the probabilities belonging to these transactions, thus obtaining a more decisive result.
This can be calculated by the naive Bayes classifier method~\cite {naivbayes}.
\begin {table} [!h] \centering \caption {The transactions of a single user (tx) assign probabilities to the clients (IP addresses), which shows the likelihood that the client is the originator of the transaction. $\mathbb {P} \left( \mathrm {IP}_i | \mathrm {tx}_j \right)$ denotes the probability that $\mathrm {IP}_i$ address created the $\mathrm {tx}_k$ transaction} \label {ipusertable} \resizebox {.45\textwidth} {!} {\begin {tabular} {llllll} \\[-1mm] \hline
 ~ & $\textrm {IP}_1$ & $\cdots$ & $\textrm {IP}_i$ & $\cdots$ & $\textrm {IP}_n$ \\ \hline
$\textrm {tx}_1$ & $\mathbb {P} \left( \textrm {IP}_1 | \textrm {tx}_1 \right)$ & ~ & $\mathbb {P} \left( \textrm {IP}_i | \textrm {tx}_1 \right)$ & ~ & $\mathbb {P} \left( \textrm {IP}_n | \textrm {tx}_1 \right)$ \\ $\textrm {tx}_2$ & $\mathbb {P} \left( \textrm {IP}_1 | \textrm {tx}_2 \right)$ & ~ & $\mathbb {P} \left( \textrm {IP}_i | \textrm {tx}_2 \right)$ & ~ & $\mathbb {P} \left( \textrm {IP}_n | \textrm {tx}_2 \right)$ \\ $\cdots$ & ~ & ~ & ~ & ~ & ~ \\ $\textrm {tx}_j$ & $\mathbb {P} \left( \textrm {IP}_1 | \textrm {tx}_j \right)$ & ~ & $\mathbb {P} \left( \textrm {IP}_i | \textrm {tx}_j \right)$ & ~ & $\mathbb {P} \left( \textrm {IP}_n | \textrm {tx}_j \right)$ \\ $\cdots$ & ~ & ~ & ~ & ~ & ~ \\ $\textrm {tx}_m$ & $\mathbb {P} \left( \textrm {IP}_1 | \textrm {tx}_m \right)$ & ~ & $\mathbb {P} \left( \textrm {IP}_i | \textrm {tx}_m \right)$ & ~ & $\mathbb {P} \left( \textrm {IP}_n | \textrm {tx}_m \right)$ \\ \hline \end {tabular}} \end {table}
Table~\ref {ipusertable} shows the transactions (denoted by tx) created by a single user.
The transactions assign probabilities to the clients (IP addresses), which indicate the likelihood that the client is the originator of the transaction.

If the ratio of the connected clients is small, the individual probabilities in the table are also low.
The probabilities of an IP address related to the different transactions can be combined by the naive Bayes classification, resulting a row of combined probabilities.
This shows how likely the IP addresses belong to the examined user.

The IP addresses will be divided into two classes, to the ``originator'' and the ``non-originator'' classes.
For each transaction, there can be at most one IP address in the originator class.
On the other hand, as a user can use multiple IP addresses to create Bitcoin transactions, after combining multiple transactions, more than one IP address can be in the originator class in the final result.

It is assumed that the Bitcoin users can be identified by a limited number of IP addresses they use when connected to the Bitcoin network.
This involves that the users do not use TOR (``The Onion Router''), proxy servers or other similar systems hiding their IP addresses.
If this does not hold, i.e.~the users use TOR, the probabilities would be distributed among several IP addresses thus resulting in small final probabilities.
We note, that the invalidity of this assumption for some users does not result in false IP address -- user pairings: only those users will be identified whom the assumption holds for.
Furthermore, previous work showed that the usage of the TOR network can be prevented by an active malicious attacker by connecting to the TOR network as well and sending malformed Bitcoin messages via the TOR exit nodes~\cite{ujot,biryukovtor}.
This kind of attack would result in users being unable to connect to the Bitcoin network via TOR.
In the current work however, we limit our analysis to regular users, i.e.~who connect to the Bitcoin network using only a few IP addresses.

By the application of the naive Bayes classifier (see Appendix~\ref {bayes_calc} for the detailed derivation), the combined probability of an IP address (IP$_i$) belonging to the $C_o$ originator class is given by
\begin {widetext} \begin {equation} \mathbb {P} \left( \textrm {IP}_i \in C_o | \mathrm {\mathbf {tx}} \right) = \frac {1} {1 + \expe \left[ \left( 1 - m \right) \ln \left( \overline {\left| A \right|} - 1 \right) + \sum\limits_{k = 1}^m \ln \left( \frac {1} {\mathbb {P} \left( \textrm {IP}_i \in C_o | \mathrm {tx}_k \right)} - 1 \right) \right]} \end {equation} \end {widetext}
where $\mathrm {\mathbf{tx}}$ denotes the vector of all considered transactions, $\overline {\left| A \right|}$ is the average of the total number of active clients through the transactions\footnote {The $\left| A \right|$ number of active clients varies through the transactions as they occur in different times. Thus, the $\overline {\left| A \right|}$ average of the different $\left| A \right|$ values is used as it is suggested in~\cite {bayes}.}
and $m$ is the number of transactions.

We note that the naive Bayes classification can only be applied if the transactions provide conditionally independent probabilities.
Otherwise the dependencies between the transactions should be determined~\cite {dependentbayes}.

\section {Data Collection}
\label {sec_datacoll}

During the data collection campaign, we used our modified Bitcoin clients to connect to the network and monitor information about transactions relayed by connected clients.
As the program code is open-source, it was straightforward to implement a monitoring client.

Our monitoring clients logged the incoming ``INV'' messages along with the IP address of the sender client and the time of reception.
These messages contain the $128$-bit hash code of the transactions which are relayed.

Using this hash code, the Bitcoin addresses, the amount of Bitcoin sent and other information of interest can then be looked up in the blockchain.

In order to monitor as large part of the Bitcoin network as possible, the modified Bitcoin clients were installed simultaneously to $140$ computers located at different parts of the world, and all of these were recording the observed traffic during the campaign.
Bitcoin clients behind firewalls usually do not allow incoming connections, i.e.~our monitoring clients can not establish connections to them.
By using a large number of monitoring clients, it is more likely that Bitcoin clients behind firewalls initiate connections to some of our monitoring clients when they enter the network.
We installed the monitoring clients on computers that are part of PlanetLab, a system maintained for network communication research.~\cite {planetlab}

The data collection campaign took slightly more than two months between $10/14/2013$ and $12/20/2013$. During this period $300$ million records were obtained, in which $4\,155\,387$ transactions and $124\,498$ IP-addresses were identified.
The collected data was imported into an SQL database server.

To calculate the probabilities described above, the total number of active clients need to be determined.
From the Bitnodes.io database~\cite {getaddr} one can look up the number of active IP addresses of the Bitcoin clients as a function of time.

\section {Results}
\label {sec_results}

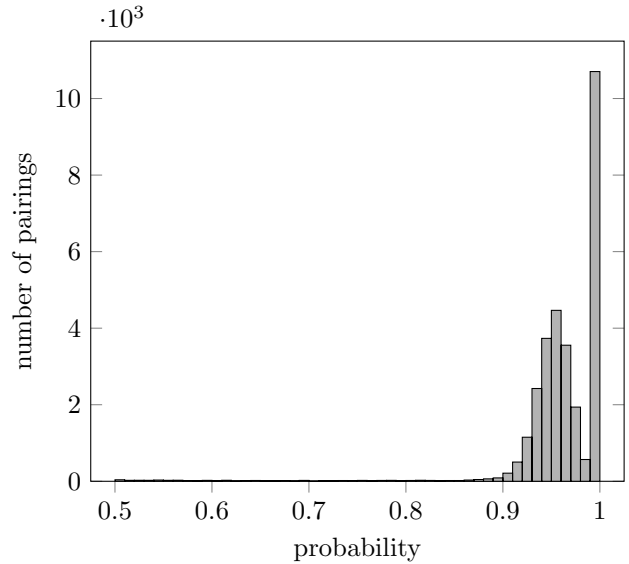
\begin {figure} [htb]
  \begin {tikzpicture}
    \begin {axis} [ybar, bar width=.01, xlabel=probability, ylabel=number of pairings, xmin=0.475, xmax=1.025, ymin=0, ymax=11500, width=.5\textwidth, xtick={0.5,0.6,0.7,0.8,0.9,1}, xtick pos=left, scaled y ticks=base 10:-3]
      \addplot+ [black,fill=black!30!white] table [] {probability_hist.txt};
    \end{axis}
  \end{tikzpicture}
  \caption {Distribution of the Probabilities Assigned to the Accepted User -- IP Address Pairings}
  \label {probab_hist}
\end {figure}

When calculating the combined probability of each IP address belonging to the specific user, the question arises when should a pairing be \emph {accepted}?
As more than one IP address can be used by each user and one IP address can be used by several users, no restriction is made of this kind.
A pairing is accepted, if its probability is higher than $0.5$.
This means that the IP address of interest has at least $0.5$ probability of being used by the user.
Figure~\ref {probab_hist} shows the distribution of the probabilities of the accepted pairings.
It can be seen, that the vast majority of the probabilities are above $0.9$, the avarage value of the pairings is $95.52\%$, so we expect the false positive rate of the results to be low.
To the best of our knowledge, the best deanonymization attempt so far could achieve a success rate of at most $0.34$~\cite {ujot}.
Two peaks can be observed on the figure, one with a maximum at $0.952$, and another close to $1$.
The first peak is due to usual clients that initiate a relatively small number of transactions.
We speculate that the other peak consists of servers offering wallet services, i.e.~servers that can be used by several people thus initiating a lot of transactions (see below in more detail).
The more initiated transactions can be taken into account, the higher the probability will be that can be assigned to the pairings.

As a result, $22\,363$ users could be identified, and altogether $1\,797$ IP addresses were assigned to them.

The imbalance is caused by three outstanding IP addresses to which $20\,680$ users are assigned.
These IP addresses probably belong to Bitcoin wallet services, which can be used for creating transactions on a website without using a private computer.
Note that the incomplete grouping of Bitcoin addresses can also result in an IP address being associated with several groups of Bitcoin addresses.
These groups actually belong to the same user, but they could not be connected in the grouping algorithm.

For the remainder data, $1.14$ IP addresses belong to one user on average.
This is due to the fact that a user can use multiple IP addresses when connecting to the Bitcoin network.
The maximum number of IP addresses identified as belonging to a single user is $8$.

\subsubsection* {Calculating the Balances}

Examining the blockchain data alone allows to investigate the time evolution of user balances before, during and after the data collection campaign.

\begin {figure*} [htb] \centering
   \subfigure [Bitcoin Owned by the Identified Clients (BTC)] {\resizebox {.46\textwidth} {!} {
      \begin {tikzpicture}
         \begin {axis} [ date coordinates in = x, xticklabel = \year/\month, date ZERO = 2012-10-18, ymin = 0, xtick={2013-01-01, 2013-07-01, 2014-01-01, 2014-07-01} ]
            \addplot+ [white, mark = none, fill = black!30!white] table [] {balance_change_fill.txt};
            \addplot+ [black, mark = none] table [] {balance_change.txt};
         \end {axis}
      \end {tikzpicture}
   }}
   \quad
   \subfigure [Exchange Rate of Bitcoin (USD)] {\resizebox {.49\textwidth} {!} {
      \begin {tikzpicture}
         \begin {axis} [ date coordinates in = x, xticklabel = \year/\month, date ZERO = 2012-10-18, scaled y ticks=true, ymin = 0, xtick={2013-01-01, 2013-07-01, 2014-01-01, 2014-07-01} ]
            \addplot+ [white, mark = none, fill = black!30!white] table [] {exchange_rate_fill.txt};
            \addplot+ [black, mark = none] table [] {exchange_rate.txt};
         \end {axis}
      \end {tikzpicture}
   }}
\caption {Balances of Bitcoin users identified in our study and the Bitcoin exchange rate. The shaded area corresponds to our data collection period.} \label {balancechange} \end {figure*}
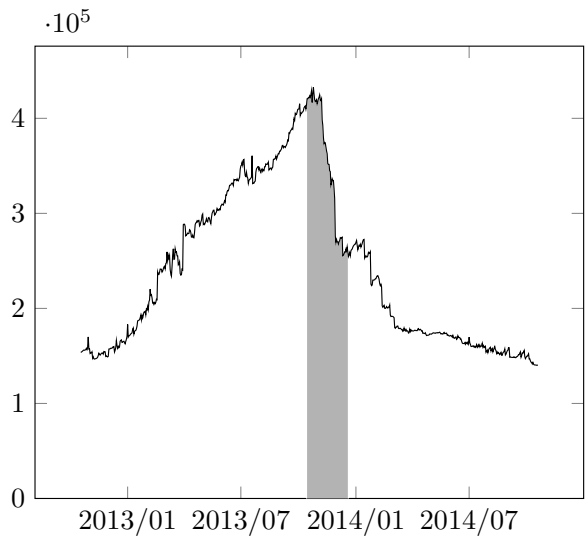
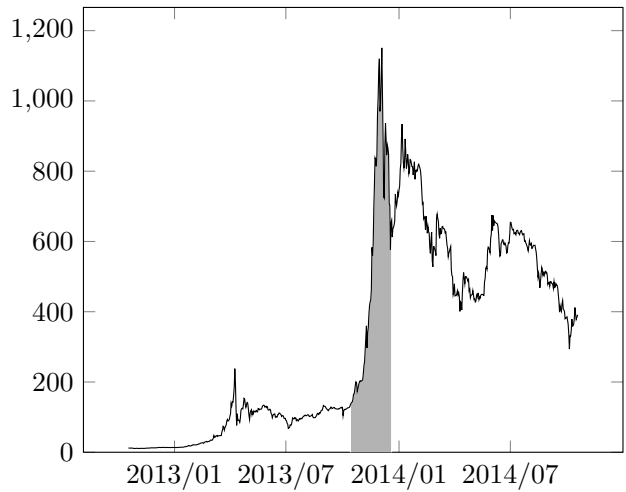

Figure~\ref {balancechange} shows the total balance of all identified users versus time.
The time interval in which the data collection was taking place is marked by a shaded area.
Before data collection, the amount of Bitcoin owned by the identified users is increasing.
This is due to the fact that some of the identified Bitcoin addresses were created before the beginning of the measurement campaign.
After the measurement, some of the identified Bitcoin addresses were not used anymore, and other new unidentified Bitcoin addresses took over their place.
The steep drop during the measurement is probably due to the significant increase of exchange rate in this time interval, which probably inspired the users to sell their Bitcoins for traditional currencies.
We found a significant, $-0.91$ linear correlation coefficient between the total amount of Bitcoin owned by the identified users and the exchange rate during the measurement period.

The total number of Bitcoins in use is constantly increasing as time goes by.
At the time of the measurement $\sim13\,500\,000$ Bitcoins were in circulation.
The amount of Bitcoins owned by the identified users reached a maximum of $432\,666$ on $10/25/2013$, which corresponds to $\sim3.2\%$ of the total amount of Bitcoins.
We believe that this ratio is a statistically representative sample, if the data were collected with uniform random sampling.
However, systematic differences could have affected the data collection as users in different parts of the world, with different intentions and technical backgrounds were possibly operating differently in the network.
The users could be protected by firewalls, thus banning incoming connections, and they could also obscure their operation by using VPN, proxy service or TOR.

\subsubsection* {Geographical Distribution of Bitcoin and the Cash Flow}

The location of IP addresses can be determined from publicly available databases such as MaxMind~\cite {maxmind}, which contains approximate locations of the IP addresses.
If the Bitcoin users use additional tools to hide their IP addresses, or if the IP addresses are located at other positions than they are registered to, the database gives false location results.
However, these inaccuracies are not relevant in the vast majority of the cases.

\begin {figure*} [p] \centering \subfigure {\includegraphics [width=.74\textwidth] {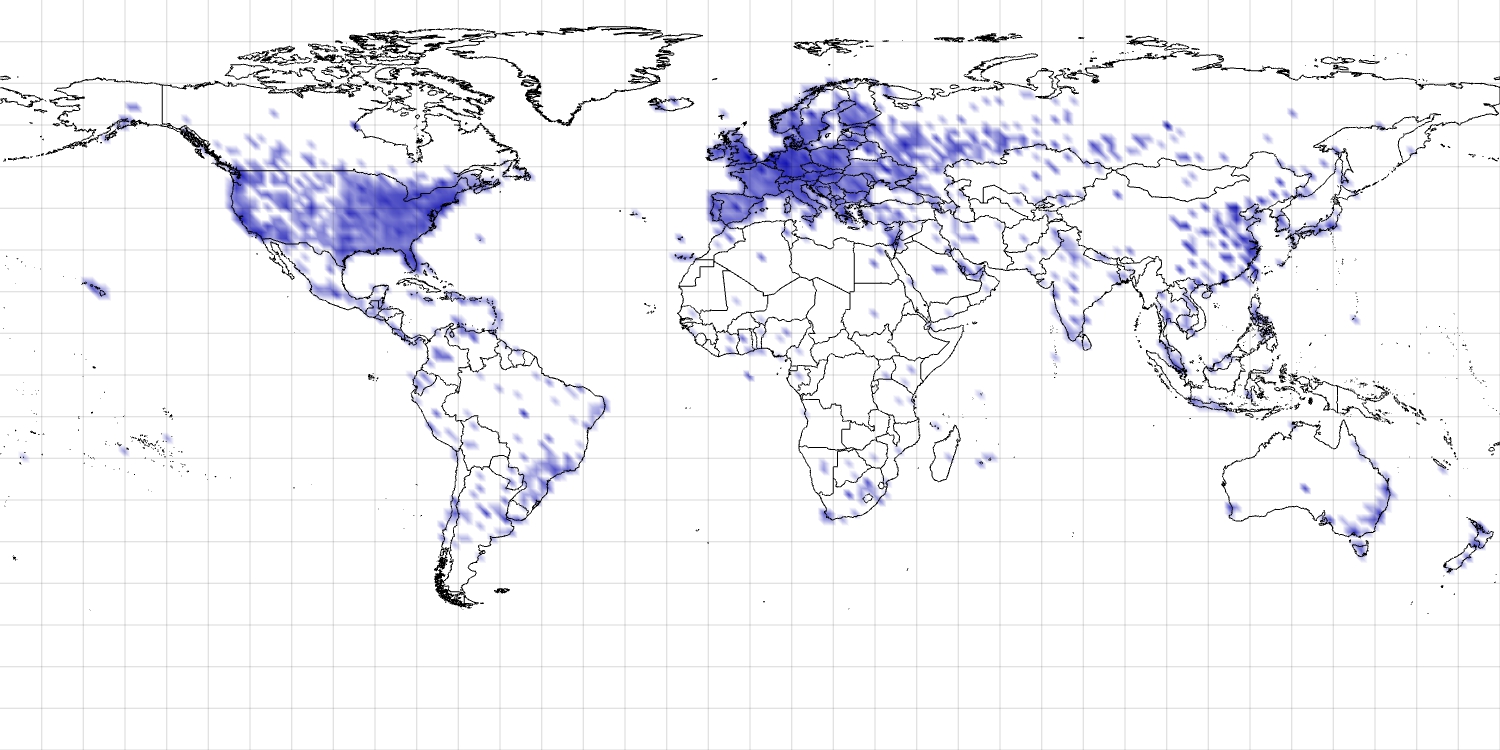}} \quad \subfigure {\includegraphics [height=.35\textwidth] {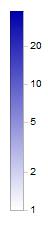}} \caption {Distribution of the Identified Bitcoin Clients ($1/100 \, \textrm {km}^2$)} \label {clients} \end {figure*}

Figure~\ref {clients} shows the distribution of the identified Bitcoin clients.
The coloring represents the logarithmic value of the density.
The identified Bitcoin clients are mostly located on the more developed regions of the world.
Note that in some countries, such as Russia or China, the Internet is regulated, therefore some interference of the connected clients (and their messages) can occur.

\begin {figure*} [p] \centering \subfigure {\includegraphics [width=.74\textwidth] {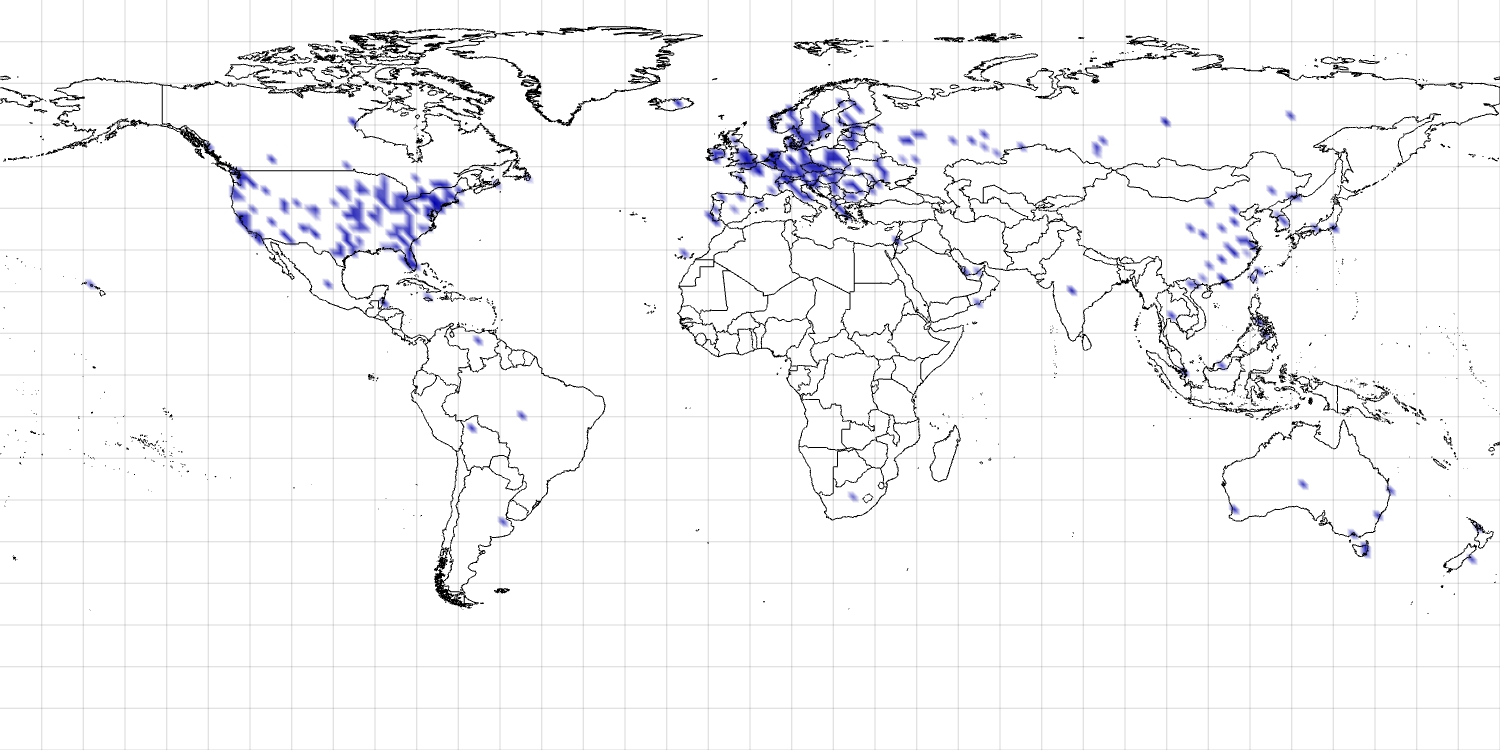}} \quad \subfigure {\includegraphics [height=.35\textwidth] {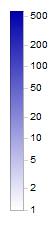}} \caption {Distribution of Bitcoin Owned by the Identified Clients on $12/20/2013$ ($1/100 \, \textrm {km}^2$)} \label {btcdistr} \end {figure*}

By the localization of the IP addresses, the geographical distribution of Bitcoin can also be determined (Figure~\ref {btcdistr}).
This figure only shows the distribution of the Bitcoins that are owned by the identified clients; the coloring is logarithmic.
The snapshot belongs to the end of the data collection period, $12/20/2013$.

The analysis detailed in Section~\ref {sec_bayesian} results in a data set of transactions and identified originators.
It is worth to examine if some originator addresses can be mapped to receiver Bitcoin addresses as well.
There are $68\,973$ transactions in which both sides could be found, and altogether $196\,971$ Bitcoins were transferred in the identified transactions.
In these transactions $7\,372$ users appear as senders and $6\,170$ appear as receivers.

\begin {figure*} [p] \centering \subfigure {\includegraphics [width=.74\textwidth] {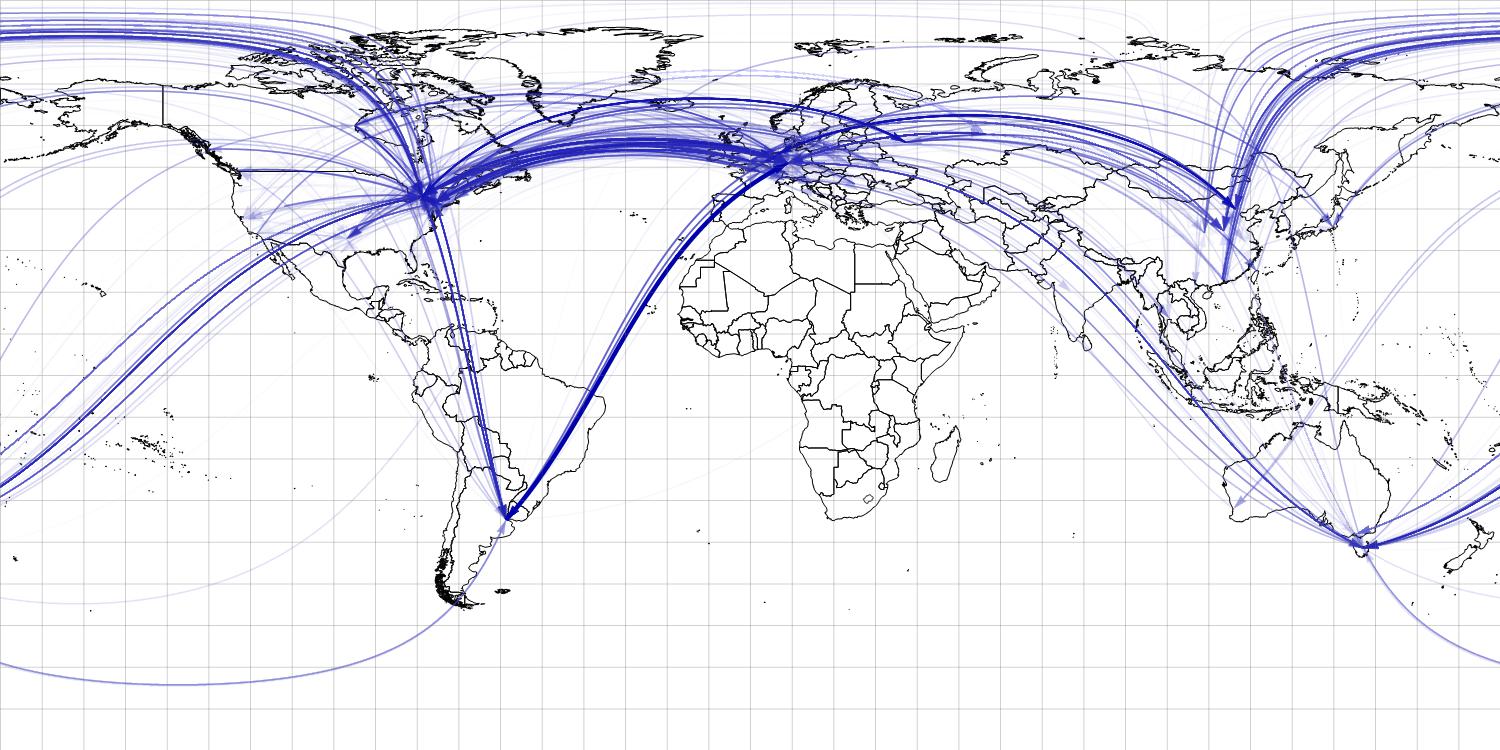}} \quad \subfigure {\includegraphics [height=.35\textwidth] {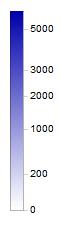}} \caption {The Flow of Bitcoin} \label {mapflow} \end {figure*}

The transactions are visualized on a world map (Figure~\ref {mapflow}).
The thickness, opacity and saturation of the arrows express the amount of Bitcoin transferred in the related transaction.
The time course of the transactions is demonstrated on a video that can be found at \emph {www.vo.elte.hu/papers/2017/bitcoin/\\supplementary/transactions\_video.mp4}.

\begin {figure} [h] \centering \includegraphics [width=.48\textwidth] {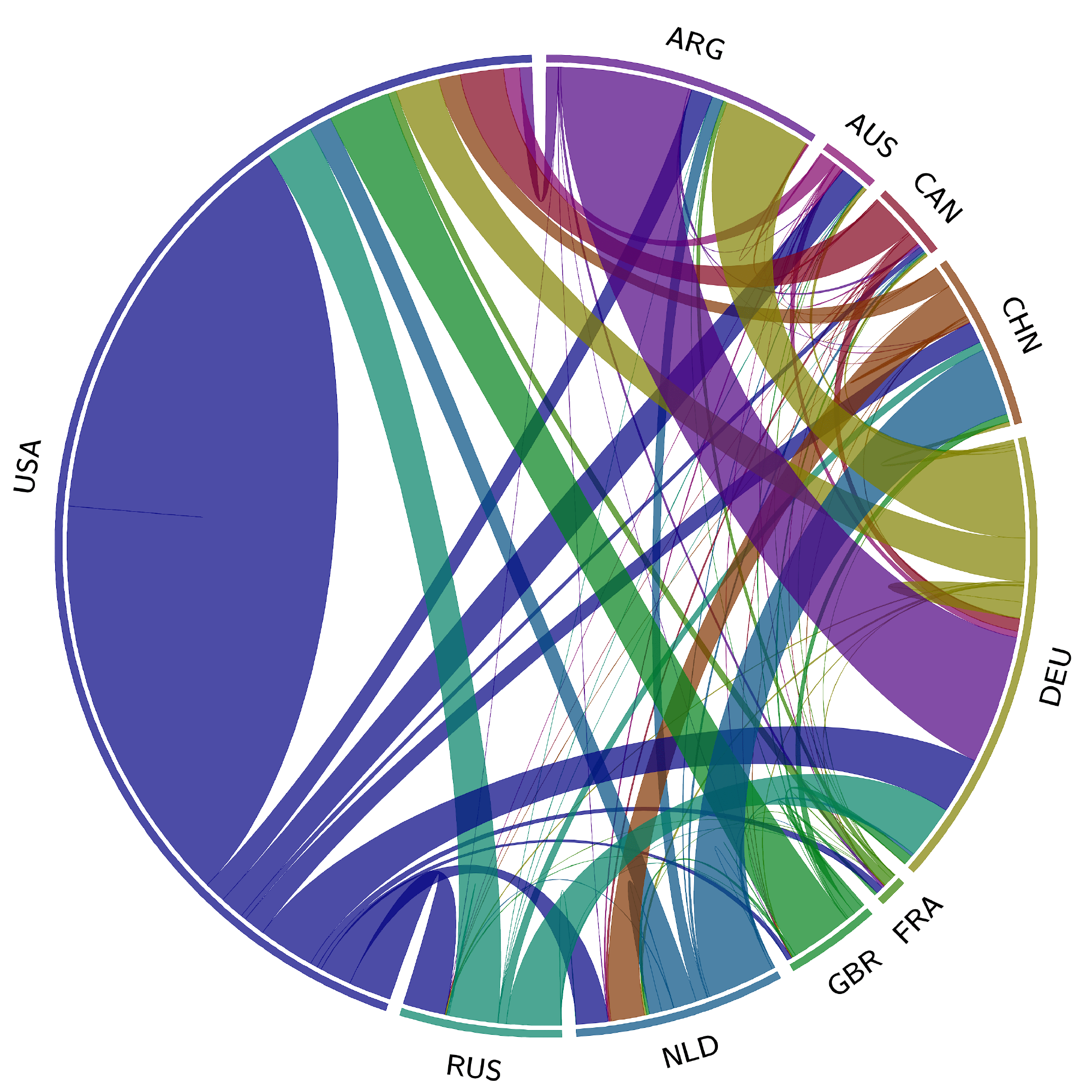} \caption {The Flow of Bitcoin Between Countries} \label {circos} \end {figure}

Let us have a look at the flow of Bitcoin between the different countries, which is illustrated in Figure~\ref {circos}.
As the vast majority of the identified Bitcoin transactions belongs to a few countries, only the top ten most significant ones are shown in the figure.
$87.5\%$ of the Bitcoins in our data set were transferred between these countries.

The different countries are indicated by arcs on the perimeter of the figure.
The colors of the links are identical with the color of the country where the Bitcoins were sent from.
A high amount of the Bitcoins ($24\,250$ Bitcoins, more than $12.3\%$ of the total amount) are transacted internally in the United States.
There are several interesting connections: the second largest flow is between Germany and Argentina ($25\,508$ Bitcoins, $13.0\%$ of the total amount), and there is a significant Bitcoin flow between China and the Netherlands as well.

\section {Related Work}
\label{sec_related}

There have been several works discussing the anonymity concerns of Bitcoin.
All of them show that the statistical processing of a huge amount of seemingly insignificant information can take the attacker closer to reveal the identity of people using Bitcoin.

Androulaki, et al.~\cite{ujketto} evaluated the privacy of Bitcoin by analyzing the system using a simulator.
After grouping Bitcoin addresses they used behavior-based clustering techniques (K-Means and Hierarchical Agglomerative Clustering algorithms) to bind the Bitcoin addresses to real users.

Reid and Harrigan~\cite{ujharom} used mainly offline data processing of the blockchain to analyze the transaction graph.
They identified its clusters and components, and analyzed the degree distribution of the user network.
They also showed that the analysis of publicly available data from social websites and forums can also reveal the Bitcoin addresses of some users, similary to previous work in different contexts~\cite{ujnegy}.

Biryukov, Khovratovich and Pustogarov provided a method which connects the users to IP addresses~\cite{ujot}.
They connected to all publicly available Bitcoin nodes (servers) and listened the messages they were relaying.
They used Bitcoin's peer discovery mechanism to link transactions to their originators even if these do not accept incoming connections: the servers that broadcast the newly connected clients' IP addresses were assumed to be the same set of servers which first relayed their transactions.
The difficulty of this method is that a lot of connections have to be established to reach good results as the number of the servers increases.
On the other hand, it promises results for Bitcoin clients which monitoring nodes cannot directly connect to (e.g.~because they are protected by firewalls and connect to a limited).
In contrast, our methodology requires direct connections to the originators and we thrive to achieve this by running a lot of Bitcoin clients accepting a large number of incoming connections.
While they use a fixed number of message relays to infer the local network of the originator, we use a short initial time span for message broadcasts to infer the actual originator.
A further main difference in our methodology is combining information from many transactions and linking addresses based on the blockchain to provide more transactions per Bitcoin user in that step.
Our probabilistic approach could be combined with the methodology in~\cite{ujot} in order to identify the ``hidden'' Bitcoin nodes with higher probability.
This also allows linking Bitcoin address groups belonging to the same user based on them being originated from the same client, even if these addresses would not be possible to link based only on the blockchain.

Koshy et al. also monitored the messages about the transactions and they classified the transactions to distinct relay patterns~\cite{ujhat}.
After applying heuristics to determine the possible owner IP addresses of the transaction, they computed simple aggregate statistics to filter out the correct Bitcoin address -- IP address pairings for both input and output addresses.

Venkatakrishnan et al. proposed a new message relay mechanism called \emph{dandelion}, which could prevent the nodes to be deanonymized with a high probability~\cite{dandelion}.
They proposed that the message propagation should have two phases: first, the message is sent to exactly one randomly chosen connected client for a random number of hops by every client, and after the first phase the message could be further broadcast with a Poisson process from the nodes received the transaction.
The authors also highlighted that requirements of the high level of anonimity and low latency are properties that can only be improved at each others expense.

Basically the following common methods are used to reveal the identities of Bitcoin users:
\begin{enumerate}
\item{analysis of transactions with multiple input and grouping the input Bitcoin addresses of the same transactions;}
\item{analysis of Bitcoin flow in the transaction graph using clustering techniques;}
\item{analysis of propagating network-layer information to bind their content to the users,}
\item{and finally using publicly available information (e.g.~in forums) to connect Bitcoin addresses to identities.}
\end{enumerate}
The methodology presented in our work combines the $1$ and $3$ types of approaches, mainly based on statistical processing of network propagation properties.

\pagebreak
\section{Conclusions}

In this paper we examined the problem of user identification in the Bitcoin network.
While Bitcoin provides a significant level of anonymity as Bitcoin addresses can be generated freely and without providing any form of personal identification, the requirement to announce new transactions on the peer to peer network opens up the possibility of linking Bitcoin addresses to the IP addresses of clients.
Our main goal was evaluating the feasibility of this procedure.

We installed a modified Bitcoin client program on over a hundred computers, which recorded the propagating messages on the network that announced new transactions.
Based on the information propagation properties of these messages, we developed a mathematical model using naive Bayes classifier method to assign Bitcoin addresses to the clients that most probably control them.
As a result, Bitcoin address -- IP address mappings were identified.
Through the IP addresses of the clients, we could determine their geographical location, which enabled the spatial analysis of distribution and flow of Bitcoin.

The method is cheap in terms of resources, the used algorithms are relatively easy to implement and can be combined with other Bitcoin-transaction related information.

All monitoring clients behaved as regular Bitcoin clients during the measurement.
Although they did not generate any transactions, the source code can be modified to do so if a better concealment is required.
Furthermore, the monitoring clients do not need to be connected to other Bitcoin users in any detectable way (i.e.~communication among them is trivially achieved outside the Bitcoin protocol), making it virtually impossible to reveal their monitoring activity.
This raises the question if the Bitcoin network might already be monitored by a similar methodology.
It can be implied that Bitcoin users should take further steps to adequately disguise their real IP addresses and preserve their anonymity.

\section* {Acknowledgements}

This work has been accomplished at the Department of Physics of Complex Systems, E\"otv\"os Lor\'and University.

We thank the PlanetLab community~\cite{planetlab} that the servers of their network could be used for the data collection.

We are also grateful to MaxMind~\cite{maxmind} for the database that could be used for the localization of the IP addresses.

\section* {Additional Information}

\paragraph {Author Contributions}
PJ analyzed the data, derived the statistical formulation and drafted the manuscript.
JS performed data collection, helped in deriving the statistical formulation and helped writing the manuscript.
DK provided software for data collection, helped with the data collection and helped write the manuscript.
GV participated in designing the study.
All authors have read the final manuscript and approved it for publication.

\paragraph {Data Availability}
All data collected during the study is made publicly available at \emph {http://www.vo.elte.hu/papers/2017/bitcoin/}.

\paragraph {Ethics Statement}
All data used in the analysis is made publicly available by the Bitcoin users as it is required by the Bitcoin protocol.
Collecting data on the level of network traffic possibly allows linking Bitcoin addresses to the IP addresses of Bitcoin users.
No other personally identifiable information beside IP address was collected about users, and no attempt was made to link IP addresses to actual people beside establishing coarse-grained geographic location.
In the shared data, IP addresses were replaced with random identifiers to prevent connecting the transactions with individuals based on other IP address related information.

\paragraph {Competing Interests}
The authors declare that they have no competing interests.

\clearpage
\section* {Bibliographies of Authors}

\begin {wrapfigure} {l} [-2mm] {2.2cm} \vspace {-4mm} \includegraphics [width=2.5cm, height=3cm, keepaspectratio] {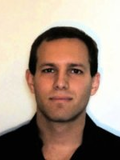} \vspace {-4mm} \end {wrapfigure} \par
\paragraph {P\'eter L Juh\'asz} obtained his MSc degree in physics in $2016$ at Budapest University of Technology and Economics, Budapest, Hungary.
After spending a semester in Delft University of Technology he has been working for Ericsson Telecommunications Hungary developing the ``Smart Services Router'' project.
He is involved in stochastic processes, cryptography and data communication networks.

\begin {wrapfigure} {l} [-2mm] {2.2cm} \vspace {-3mm} \includegraphics [width=2.5cm, height=3cm, keepaspectratio] {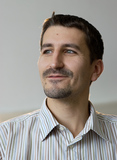} \vspace {-4mm} \end {wrapfigure} \par
\paragraph {J\'ozsef St\'eger} received his PhD in physics in $2010$ at E\"otv\"os Lor\'and University in Budapest, Hungary.
He is a founder member of the Communication Networks Laboratory at E\"otv\"os Lor\'and University since $2000$.
Earlier, his research focused on computer communication networks, especially the active measurement of network delay and the inference of delay statistics of the internal hidden part of the network.
He took part in numerous domestic and international research projects in this field.
Lately, he turned interest to the investigation of motion patterns present in ballgames and the localization and characterization of acoustic events.

\begin {wrapfigure} [8] {l} [-2mm] {2.2cm} \vspace {-4mm} \includegraphics [width=2.2cm, height=3cm, keepaspectratio] {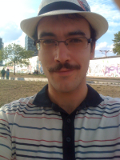} \vspace {-4mm} \end {wrapfigure} \par
\paragraph {D\'aniel Kondor} obtained his PhD in physics in $2015$ at E\"otv\"os Lor\'and University in Budapest, Hungary.
He has been working as a postdoctoral researcher at the Senseable City Lab at the Massachusetts Institute of Technology since then.
His research has focused on complex social, communication and economic networks and utilizing novel large-scale data collection technologies to gain better understanding of social phenomena; this includes data from mobile network usage logs, social networks like Twitter or the Bitcoin cryptocurrency.
During his PhD, he also worked as an intern at Ericsson Research as part of their academic collaboration program.

\begin {wrapfigure} [8] {l} [-2mm] {2.2cm} \vspace {-4mm} \includegraphics [width=2.2cm, height=3cm, keepaspectratio] {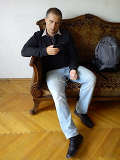} \vspace {-4mm} \end {wrapfigure} \par
\paragraph {G\'abor Vattay} (Full Professor, E\"otv\"os Lor\'and University, Budapest) received his PhD in physics from the Hungarian Academy of Sciences in $1994$.
His main interest is complex adaptive systems, social networks, nonlinear dynamics and chaos.
He has published over $100$ papers in physics and networking and edited the book Complex Dynamics in Communication Networks.

\newgeometry{margin=0.15\paperwidth}
\onecolumn

\begin{appendices}
\FloatBarrier

\clearpage
\
\section* {Appendix - Derivation of Naive Bayes Classifier Method}
\label {bayes_calc}

The model classifies the clients into the \emph{originator} and \emph{non-originator} classes ($C_o$ and $C_n$ respectively) based on their IP addresses and by considering $m$ transactions.
Transactions are denoted by $\mathrm {\mathbf {tx}} = \{ \mathrm {tx}_i \}$, ($i \in \left[ 1;\, m \right]$).

Consider a single IP address, and let us examine the probabilities that the different transactions assign to it.
Using the Bayes theorem the probability of belonging to the originator class is
\begin {equation} \mathbb {P} \left( \textrm {IP}_i \in C_o | \mathrm {\mathbf {tx}} \right) = \frac {\mathbb {P} \left( \textrm {IP}_i \in C_o \right)} {\mathbb {P} \left( \mathrm {\mathbf {tx}} \right)} \, \mathbb {P} \left( \mathrm {\mathbf {tx}} | \textrm {IP}_i \in C_o \right) \end {equation}
where $\mathbb {P} \left( \textrm {IP}_i \in C_o \right)$ is the frequency of $C_o$ class (a priori probability).
By assuming that the probabilities $\mathbb {P} \left( \mathrm {\mathbf {tx}} | \textrm {IP}_i \in C_o \right)$ are conditionally independent, the expression can be simplified.
\begin {equation} \mathbb {P} \left( \textrm {IP}_i \in C_o | \mathrm {\mathbf {tx}} \right) = \frac {\mathbb {P} \left( \textrm {IP}_i \in C_o \right)} {\mathbb {P} \left( \mathrm {\mathbf {tx}} \right)} \prod\limits_{i = 1}^m \mathbb {P} \left( \mathrm {tx}_i | \textrm {IP}_i \in C_o \right) \end {equation}
Bayes theorem can be applied again to the factors in the product.
\begin {equation} \mathbb {P} \left( \textrm {IP}_i \in C_o | \mathrm {\mathbf {tx}} \right) = \frac {\mathbb {P} \left( \textrm {IP}_i \in C_o \right)} {\mathbb {P} \left( \mathrm {\mathbf {tx}} \right)} \prod\limits_{i = 1}^m \frac {\mathbb {P} \left( \textrm {IP}_i \in C_o | \mathrm {tx}_i \right) \mathbb {P} \left( \mathrm {tx}_i \right)} {\mathbb {P} \left( \textrm {IP}_i \in C_o \right)} = \underbrace {\frac {\prod\limits_{i = 1}^m \mathbb {P} \left( \mathrm {tx}_i \right)} {\mathbb {P} \left( \mathrm {\mathbf {tx}} \right)}}_{\mathrm {const}} \cdot \frac {\prod \limits_{i = 1}^m \mathbb {P} \left( \textrm {IP}_i \in C_o | \mathrm {tx}_i \right)} {\mathbb {P} \left( \textrm {IP}_i \in C_o \right)^{m - 1}} \end {equation}
The first factor is constant (depends only on the data), and it can be eliminated by the normalization of the probabilities.
As $\mathbb {P} \left( \textrm {IP}_i \in C_o | \mathrm {\mathbf {tx}} \right) + \mathbb {P} \left( \textrm {IP}_i \in C_n | \mathrm {\mathbf {tx}} \right) = 1$ is valid,
\begin {equation} \begin {split} \mathbb {P} \left( \textrm {IP}_i \in C_o | \mathrm {\mathbf {tx}} \right) &= \underbrace {\frac {1} {\frac {\prod \limits_{i = 1}^m \mathbb {P} \left( \textrm {IP}_i \in C_o | \mathrm {tx}_i \right)} {\mathbb {P} \left( \textrm {IP}_i \in C_o \right)^{m - 1}} + \frac {\prod \limits_{i = 1}^m \mathbb {P} \left( \textrm {IP}_i \in C_n | \mathrm {tx}_i \right)} {\mathbb {P} \left( \textrm {IP}_i \in C_n \right)^{m - 1}}}}_{\mathrm {const}} \cdot \frac {\prod \limits_{i = 1}^m \mathbb {P} \left( \textrm {IP}_i \in C_o | \mathrm {tx}_i \right)} {\mathbb {P} \left( \textrm {IP}_i \in C_o \right)^{m - 1}} = \\ &= \frac {1} {\frac {\prod \limits_{i = 1}^m \mathbb {P} \left( \textrm {IP}_i \in C_o | \mathrm {tx}_i \right)} {\mathbb {P} \left( \textrm {IP}_i \in C_o \right)^{m - 1}} + \frac {\prod \limits_{i = 1}^m \left( 1 - \mathbb {P} \left( \textrm {IP}_i \in C_o | \mathrm {tx}_i \right) \right)} {\left( 1 - \mathbb {P} \left( \textrm {IP}_i \in C_o \right) \right)^{m - 1}}} \cdot \frac {\prod \limits_{i = 1}^m \mathbb {P} \left( \textrm {IP}_i \in C_o | \mathrm {tx}_i \right)} {\mathbb {P} \left( \textrm {IP}_i \in C_o \right)^{m - 1}} \end {split} \end {equation}
The expression can be simplified further.
\begin {equation} \begin {split} \mathbb {P} \left( \textrm {IP}_i \in C_o | \mathrm {\mathbf {tx}} \right) &= \frac {\frac {\prod\limits_{k = 1}^m \mathbb {P} \left( \textrm {IP}_i \in C_o | \mathrm {tx}_k \right)} {\mathbb {P} \left( \textrm {IP}_i \in C_o \right)^{m - 1}}} {\frac {\prod\limits_{k = 1}^m \mathbb {P} \left( \textrm {IP}_i \in C_o | \mathrm {tx}_k \right)} {\mathbb {P} \left( \textrm {IP}_i \in C_o \right)^{m - 1}} + \frac {\prod\limits_{k = 1}^m \left( 1 - \mathbb {P} \left( \textrm {IP}_i \in C_o | \mathrm {tx}_k \right) \right)} {\left( 1 - \mathbb {P} \left( \textrm {IP}_i \in C_o \right) \right)^{m - 1}}} = \\ &= \frac {1} {1 + \frac {\mathbb {P} \left( \textrm {IP}_i \in C_o \right)^{m - 1}} {\left( 1 - \mathbb {P} \left( \textrm {IP}_i \in C_o \right) \right)^{m - 1}} \cdot \frac {\prod\limits_{k = 1}^m \left( 1 - \mathbb {P} \left( \textrm {IP}_i \in C_o | \mathrm {tx}_k \right) \right)} {\prod\limits_{k = 1}^m \mathbb {P} \left( \textrm {IP}_i \in C_o | \mathrm {tx}_k \right)}} \end {split} \end {equation}
$\mathbb {P} \left( \textrm {IP}_i \in C_o \right)$ is the initial frequency of occurrence of the clients in the $C_o$ class, which is $1 / \left| A \right|$.
Although a Bitcoin client can use multiple IP addresses in the network, it is assumed that the $1 / \left| A \right|$ value is a good approximation for the initial frequency in the vast majority of the cases.
The total number of active clients varies with time in the scale of all considered transactions.
Thus, the $\overline {\left| A \right|}$ average of the different $\left| A \right|$ values is used as suggested in~\cite {bayes}.
\begin {equation} \begin {split} \mathbb {P} \left( \textrm {IP}_i \in C_o | \mathrm {\mathbf{tx}} \right) &= \frac {1} {1 + \left( \frac {\frac {1} {\overline {\left| A \right|}}} {1 - \frac {1} {\overline {\left| A \right|}}} \right)^{m - 1} \cdot \prod\limits_{k = 1}^m \left( \frac {1} {\mathbb {P} \left( \textrm {IP}_i \in C_o | \mathrm {tx}_k \right)} - 1 \right)} = \\ &= \frac {1} {1 + \left( \overline {\left| A \right|} - 1 \right)^{1 - m} \cdot \prod\limits_{k = 1}^m \left( \frac {1} {\mathbb {P} \left( \textrm {IP}_i \in C_o | \mathrm {tx}_k \right)} - 1 \right)}, \end {split} \end {equation}

This formula brings in a technical problem. 
Huge numbers are multiplied together in the product, which become significantly biased in regular number representations by rounding and may result in overflow.
To relax this problem, the second term of the denominator is written in an exponential form.
\begin {equation} \begin {split} &\mathbb {P} \left( \textrm {IP}_i \in C_o | \mathrm {\mathbf{tx}} \right) = \frac {1} {1 + \mathrm {e}^\xi}, \\ &\xi = \left( 1 - m \right) \ln \left( \overline {\left| A \right|} - 1 \right) + \sum\limits_{k = 1}^m \ln \left( \frac {1} {\mathbb {P} \left( \textrm {IP}_i \in C_o | \mathrm {tx}_k \right)} - 1 \right). \end {split} \end {equation}
This results in the following practical formula.
\begin {equation} \mathbb {P} \left( \textrm {IP}_i \in C_o | \mathrm {\mathbf{tx}} \right) = \frac {1} {1 + \expe \left[ \left( 1 - m \right) \ln \left( \overline {\left| A \right|} - 1 \right) + \sum\limits_{k = 1}^m \ln \left( \frac {1} {\mathbb {P} \left( \textrm {IP}_i \in C_o | \mathrm {tx}_k \right)} - 1 \right) \right]} \end {equation}

This formula enables us to combine the probabilities assigned to the IP addresses by the transactions.

\end{appendices}

\end{document}